\documentclass{PoS}

\title{Hadron masses from fixed topology simulations: parity partners and SU(2) Yang-Mills results}

\ShortTitle{Hadron masses from fixed topology simulations}

\author{\speaker{Arthur Dromard}, Christopher Czaban, Marc Wagner \\
        Goethe-Universit\"at Frankfurt am Main \\
        Institut f\"ur Theoretische Physik \\
        Max-von-Laue-Stra\ss e 1, D-60438 Frankfurt am Main, Germany \\
        E-mail: \email{dromard@th.physik.uni-frankfurt.de}, \email{czaban@th.physik.uni-frankfurt.de}, \email{mwagner@th.physik.uni-frankfurt.de}}

\abstract{Lattice QCD simulations tend to get stuck in a single topological sector at fine lattice spacing, or when using chirally symmetric quarks. In such cases computed observables differ from their full QCD counterparts by finite size effects, which need to be understood on a quantitative level. We discuss extensions of existing relations from the literature between correlation functions at fixed topology and hadron masses at unfixed topology. Particular focus is put on disentangling positive and negative parity states, which mix, when the topological charge is fixed. We also present numerical results for SU(2) Yang-Mills Theory.}

\FullConference{The 32nd International Symposium on Lattice Field Theory,\\
		23-28 June, 2014\\
		Columbia University New York, NY}

\begin{document}

\section{Introduction}

Topology freezing or fixing are important issues in quantum field theory, in particular in QCD. For example Monte Carlo simulations with a local update algorithm tend to be stuck in a single topological sector at lattice spacings $a \lesssim 0.05 \, \textrm{fm}$, which are nowadays still rather fine, but realistic \cite{Luscher:2011kk}. Similarly, when simulating chirally symmetric overlap quarks, the corresponding algorithms are not able to generate transitions between different topological sectors (cf.\ e.g.\ \cite{Aoki:2008tq}).

In view of these issues it is important to develop methods, which allow us to obtain physically meaningful results (i.e.\ results corresponding to unfixed topology) from fixed topology simulations. The starting point for our work are the seminal papers \cite{Brower:2003yx,Aoki:2007ka}. The calculations from these papers have been extended in \cite{Dromard:2013wja,Dromard:2014ela} by including fixed topology correction terms up to $\mathcal{O}(1/V^3)$. Tests and applications of these equations to quantum mechanics, $2d$ $O(3)$ model and the Schwinger model can be found in \cite{Bautista:2014tba,Bietenholz:2011ey,Bietenholz:2012sh,Czaban:2013haa,Dromard:2013wja,Czaban:2014gva,Dromard:2014ela,Gerber:2014bia}. Here we discuss parity mixing due to topology fixing and its consequences, when extracting hadron masses from fixed topology simulations. We also present results on SU(2) Yang-Mills theory.

\section{BCNW equation and extensions}

\subsection{\label{sub:BCNW-equation_paragraph}BCNW equation and extraction of hadron masses from fixed topology simulations}

The partition function and the two-point correlation function of a hadron creation operator $O$ at fixed topological charge $Q$ and finite spacetime volume $V$ are
\begin{equation}
\begin{aligned}
 & Z_{Q,V} = \int DA\, D\psi\, D\bar{\psi}\,\delta_{Q,Q[A]}e^{-S_{E}[A,\bar{\psi},\psi]} \\
 & C_{Q,V}(t) = \frac{1}{Z_{Q,V}}\int DA\, D\psi\, D\bar{\psi}\,\delta_{Q,Q[A]}O^{\dagger}(t)O(0)e^{-S_{E}[A,\bar{\psi},\psi]}.
\end{aligned}
\end{equation}

For large $V$ one can use a saddle point approximation and expand the correlation function \cite{Brower:2003yx},
\begin{equation}
\label{eq:BCNW} C_{Q,V}(t)=\alpha(0)\exp\bigg(-M_{H}(0)t-\frac{M_{H}^{(2)}(0)t}{2\chi_{t}V}\bigg(1-\frac{Q^{2}}{\chi_{t}V}\bigg)\bigg)+\mathcal{O}\bigg(\frac{1}{\chi_{t}^{2}V^{2}}\bigg),
\end{equation}
where $\alpha(0)=\alpha(\theta=0)$ is a constant, $M_{H}(0)=M_{H}(\theta=0)$ the physical hadron mass (i.e.\ at unfixed topology), $\theta$ denotes the QCD vacuum angle and $\chi_{t}$ the topological susceptibility. In the following we will refer to this equation as BCNW equation\footnote{BCNW stands for R.~Brower, S.~Chandrasekharan, J.~W.~Negele and U.-J.~Wiese}. In order to be a valid approximation, certain conditions have to be fulfilled, e.g.\ $1/\chi_{t}V\ll1$, $|Q|/\chi_{t}V\ll1$ and $|M_{H}^{(2)}(0)t|/\chi_{t}V\ll1$. For a detailed discussion cf.\ \cite{Dromard:2014ela}, Section~4. 

A straightforward method to determine physical hadron masses (i.e.\ at unfixed topology) from fixed topology simulations based on the BCNW equation has been proposed in \cite{Brower:2003yx}:
\begin{enumerate}
\item Perform simulations at fixed topology for different topological charges $Q$ and spacetime volumes $V$, for which the BCNW equation is a good approximation, i.e.\ where the above mentioned conditions are fulfilled. Compute $C_{Q,V}(t)$ for each simulation.

\item Determine the physical hadron mass $M_{H}(0)$, $M_{H}^{(2)}(0)$ and $\chi_{t}$ by fitting the BCNW equation (\ref{eq:BCNW}) to the numerical results for $C_{Q,V}(t)$ obtained in step 1.
\end{enumerate}

\subsection{Higher orders in $1/V$}

In the derivation of the BCNW equation (\ref{eq:BCNW}) all fixed topology corrections proportional to $1/V$ have been taken into account as well as some proportional to $1/V^2$. In \cite{Dromard:2013wja,Dromard:2014ela} we have extended this expansion by including all terms of $\mathcal{O}(1/V^2)$ and $\mathcal{O}(1/V^3)$. While there are only 4 parameters in the BCNW equation ($\alpha(0)$, $M_{H}(0)$, $M_{H}^{(2)}(0)$ and $\chi_{t}$), there are 8 and 11 parameters in the corresponding $1/V^2$ and $1/V^3$ versions, respectively. Such large numbers of unknown parameters might lead to unstable fits, when using methods to determine $M_{H}(0)$ similar to that discussed in Subsection~\ref{sub:BCNW-equation_paragraph}. As a compromise between using higher orders at the one hand and stable fits on the other hand we advocate to use the $1/V^3$ version with the 4 parameters of the BCNW equation only (the remaining 7 parameters are set to zero):
\begin{equation}
\label{eq:1} C_{Q,V}(t)=\frac{\alpha(0)}{\sqrt{1+M_{H}^{(2)}(0)t/\chi_{t}V}}\exp\bigg(-M_{H}(0)t-\frac{1}{\chi_{t}V}\bigg(\frac{1}{1+M_{H}^{(2)}(0)t/\chi_{t}V}-1\bigg)\frac{1}{2}Q^{2}\bigg) .
\end{equation}
A comparison of $M_{H}(0)$ determinations using this equation and the BCNW equation in quantum mechanics suggests that it is advantageous to use (\ref{eq:1}) (cf.\ \cite{Dromard:2014ela} for details).

\subsection{\label{sub:Parity-mixing.}Parity mixing}

Parity $P$ is not a symmetry at $\theta \neq 0$. Therefore, states at $\theta \neq 0$ cannot be classified according to parity and it is not possible to construct two-point correlation functions $\mathcal{C}_{\theta,V}(t)$, where only $P=-$ or $P=+$ states contribute. Similarly, $C_{Q,V}(t)$ contains contributions of states both with $P=-$ and $P=+$, since it is the Fourier transform of $\mathcal{C}_{\theta,V}(t)$. Consequently, one has to determine the masses of $P=-$ and $P=+$ parity partners from the same two-point correlation functions. While usually there are little problems for the lighter state (in the case of mesons typically the $P=-$ ground state), its parity partner (the $P=+$ ground state) has to be treated as an excitation. To precisely determine the mass of an excited state, a single correlator is in most cases not sufficient. For example to extract a first excitation it is common to study at least a $2\times2$ correlation matrix formed by two hadron creation operators, which generate significant overlap to both the ground state and the first excitation.

We discuss the determination of $P=-$ and $P=+$ parity partners from fixed topology computations in a simple setup, a $2\times2$ correlation matrix
\begin{equation}
C_{Q,V}(t)=
\left(\begin{array}{cc}
C_{Q,V}^{--}(t) & C_{Q,V}^{-+}(t)\\
C_{Q,V}^{+-}(t) & C_{Q,V}^{++}(t)
\end{array}\right)
\quad , \quad C_{Q,V}^{jk}(t)\equiv\frac{1}{Z_{Q,V}}\int DA\, D\psi\, D\bar{\psi}\,\delta_{Q,Q[A]}O_{j}^{\dagger}(t)O_{k}(0)e^{-S_{E}[A,\bar{\psi},\psi]}
\end{equation}
with hadron creation operators $O_{-}$ and $O_{+}$ generating at small $\theta$ mainly $P=-$ and $P=+$ states, respectively. Without loss of generality we assume that the ground state at $\theta=0$ has $P=-$, denoted by $H_{-}$, and the first excitation has $P=+$, denoted by $H_{+}$. Starting from the expression of a correlation function at fixed $\theta$, where we consider the two states $H_{-}$ and $H_{+}$,
\begin{equation}
\mathcal{C}_{\theta,V}^{jk}(t)\mathcal{Z}_{\theta,V}=\Big(\alpha_{-}^{jk}(\theta,V_{s})e^{-M_{H_{-}}(\theta)t}+\alpha_{+}^{jk}(\theta,V_{s})e^{-M_{H_{+}}(\theta)t}\Big)e^{-E_{0}(\theta,V_{s})T}
\end{equation}
with the spatial volume $V_{s}$  and the temporal extension $T$, one can derive the form of the four elements of the fixed topology correlation matrix by applying the same techniques used to derive the BCNW equation \cite{Dromard:2014ela}. Neglecting terms of $\mathcal{O}(1/V^2)$ the result is
\begin{eqnarray}
\label{EQN001} & & \hspace{-0.7cm} C_{Q,V}^{--}(t) = a_{11}e^{-M_{H_{-}}(0)t}\bigg(1-\frac{M_{H_{-}}^{(2)}(0)t}{2\chi_{t}V}\bigg)+\frac{b_{22}}{\chi_{t}V}e^{-M_{H_{+}}(0)t}\\
 & & \hspace{-0.7cm}C_{Q,V}^{-+}(t) = \frac{iQa_{12}}{\chi_{t}V}e^{-M_{H_{-}}(0)t}+\frac{iQb_{12}}{\chi_{t}V}e^{-M_{H_{+}}(0)t}\\
 & & \hspace{-0.7cm}C_{Q,V}^{+-}(t) = \frac{iQa_{21}}{\chi_{t}V}e^{-M_{H_{-}}(0)t}+\frac{iQb_{21}}{\chi_{t}V}e^{-M_{H_{+}}(0)t}\\
\label{eq:3} & & \hspace{-0.7cm} C_{Q,V}^{++}(t) = \frac{a_{22}}{\chi_{t}V}e^{-M_{H_{-}}(0)t}+b_{22}e^{-M_{H_{+}}(0)t}\bigg(1-\frac{M_{H_{-}}^{(2)}(0)t}{2\chi_{t}V}\bigg) .
\end{eqnarray}
The difficulties due to parity mixing, when trying to determine $M_{H_{+}}$, are nicely illustrated by (\ref{eq:3}): the contamination of $C_{Q,V}^{++}(t)$ by the $P=-$ state is proportional by $1/V$ (and, therefore, might be small), but the signal term is exponentially suppressed in $t$, proportional to $e^{-(M_{H_{+}}(0) - M_{H_{-}}(0)) t}$; consequently, at large $t$ the $P=-$ state will inevitably dominate. As mentioned above, a possible solution might be to determine $M_{H_{-}}(0)$ and $M_{H_{+}}(0)$ at the same time by fitting (\ref{EQN001}) to (\ref{eq:3}) to a $2 \times 2$ correlation matrix.

Of course, when one is only interested in $M_{H_{-}}$, the situation is much simpler. In particular when $M_{H_{-}} \ll M_{H_{+}}$, the BCNW equation or its improved version (\ref{eq:1}) can be used in a straightforward way as discussed in Subsection~\ref{sub:BCNW-equation_paragraph}. In the next section we will study Yang-Mills theory at fixed topology following this strategy.

\section{Computations in SU(2) Yang-Mills theory at fixed topology}

\subsection{\label{SEC001}Simulation setup}

In the continuum the SU(2) Yang-Mills Lagrangian is
\begin{equation}
\mathcal{L}(A)=\frac{1}{4g^{2}}F_{\mu\nu}^{a}F_{\mu\nu}^{a}.
\end{equation}
The corresponding lattice action we use is the standard plaquette action with $\beta=2.5$, which amounts to a lattice spacing $a \approx 0.073 \, \textrm{fm}$. We have generated gauge configurations for spacetime volumes $\hat{V}=V/a^{4}\in\{14^{4}\,,\,15^{4}\,,\,16^{4}\,,\,18^{4}\}$. For each volume the static quark-antiquark potential ${\mathcal{V}}_{q\bar{q}}(r)$ for various quark-antiquark separations $r=a,2a,\ldots,6a$ has been computed on 4000 gauge configurations. For each of these gauge configurations the topological charge has been computed using a cooling procedure explained in \cite{de Forcrand:1997sq}.

\subsection{\label{SEC387}The static potential}

\begin{figure}[htb]
\begin{center}
\includegraphics[width=12.5cm]{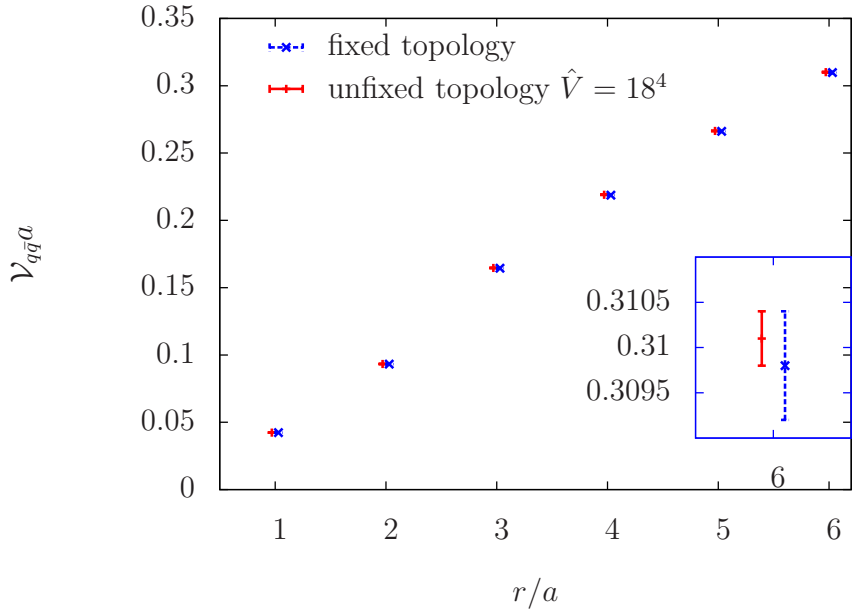}
\caption{\label{FIG003}Comparison of static potential results obtained from fixed topology Wilson loops (using (2.3) and a single combined fit) and at unfixed topology (at $\hat{V}=18^{4}$). Since unfixed and fixed topology results are identical within statistical errors, they have slightly been shifted to the left and right, respectively, for better visibility.}
\end{center}
\end{figure}

To obtain the physical static potential from Wilson loop averages, separately computed in different topological sectors $Q\in\{0\,,\,1\,,\,\ldots\,,\,7\}$ and volumes $\hat{V}\in\{14^{4}\,,\,15^{4}\,,\,16^{4}\,,\,18^{4}\}$, denoted by $\langle W_{Q,V}(r,t)\rangle$, we proceed as sketched in Subsection~\ref{sub:BCNW-equation_paragraph} and discussed in detail in Section~5.3.4 of \cite{Dromard:2014ela}.
\begin{itemize}
\item We perform $\chi^{2}$ minimizing fits of either the BCNW equation (\ref{eq:BCNW}) or the corresponding improved version (\ref{eq:1}) with respect to their parameters $\alpha(r)$, ${\mathcal{V}}_{q\bar{q}}(r)$, ${\mathcal{V}}''_{q\bar{q}}(r)$ ($r=a,2a,\ldots,6a$) and $\chi_{t}$ to the numerical results for $\langle W_{Q,V}(r,t)\rangle$.

\item We either perform a single combined fit to all considered separations $r=a,2a,\ldots,6a$ or six separate fits, one for each of the six separations. In the latter case one obtains also six different results for the topological susceptibility $\chi_{t}$.

\item Since the validity of both the BCNW equation (\ref{eq:BCNW}) and the corresponding improved version (\ref{eq:1}) requires certain conditions (cf.\ Subsection~\ref{sub:BCNW-equation_paragraph}), we include only Wilson loops $\langle W_{Q,V}(r,t)\rangle$ with $1/\chi_{t}V,|Q|/\chi_{t}V<1.0$ in the fits.
\end{itemize}
In Figure~\ref{FIG003} we compare the static potential obtained from fixed topology Wilson loops (using (\ref{eq:1}) and a single combined fit) to the static potential computed without topology fixing (at $\hat{V}=18^{4}$). There is excellent agreement within statistical errors. Qualitatively identical results have been obtained for the BCNW equation, or when performing six separate fits to the six separations.

\begin{figure}[htb]
\begin{center}
\includegraphics[width=12.5cm]{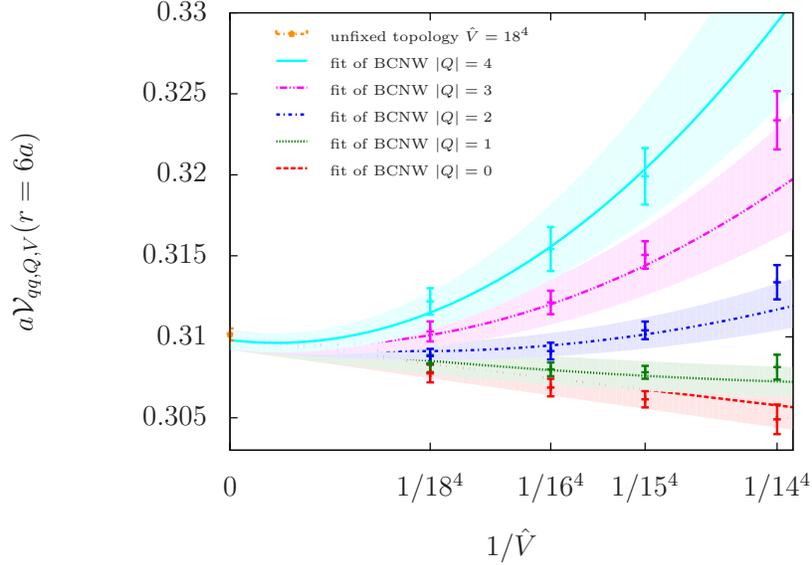}
\caption{\label{FIG690}The fixed topology static potential $\mathcal{V}_{q\bar{q},Q,V}(r=6a)$ for various $Q=0,1,\ldots4$ as a function of $1/\hat{V}$ and the
corresponding BCNW expansions.}
\vspace{-4mm}
\end{center}
\end{figure}

For $|Q|=0,1,\ldots4$ the obtained values for $\mathcal{V}_{q\bar{q},Q,V}(r=6a)$ are plotted in Figure~\ref{FIG690}. We observe a strong dependence of the static potential on the topological sector, which becomes increasingly prominent for smaller spacetime volumes. The fixed topology static potential is expected to behave as the exponent of the BCNW equation (\ref{eq:BCNW}). The corresponding curves for $Q=0,1,\ldots4$ with parameters ${\mathcal{V}}_{q\bar{q}}(r=6a)$, ${\mathcal{V}}''_{q\bar{q}}(r=6a)$ and $\chi_{t}$ determined by the previously discussed fits (using (\ref{eq:BCNW}) and a single combined fit) are also shown in Figure~\ref{FIG690}. One can clearly see that (\ref{eq:BCNW}) nicely describes the numerical results for $\mathcal{V}_{q\bar{q},Q,V}(r=6a)$.

We conclude that one can obtain a correct and accurate physical static potential (corresponding to unfixed topology) from Wilson loops separately computed in different topological sectors. 

\subsection{\label{SEC588}The topological susceptibility}

In Table~\ref{TAB003} we present results for the topological susceptibility extracted from fixed topology Wilson loops $\langle W_{Q,V}(r,t)\rangle$. As explained in the previous section we have used either the BCNW equation (\ref{eq:BCNW}) or the improved version (\ref{eq:1}) and either a single fit to all considered separations $r=a,2a,\ldots,6a$ or six different fits, one for each of the six separations. In the latter case one obtains also six different results for the topological susceptibility $\chi_{t}$.

\begin{table}[htb]
\begin{centering}

{\small{}}%
\begin{tabular}{|c|c|c|c|c|c|c|}
\hline 
{\small{method }} & {\small{$r=a$ }} & {\small{$r=2a$ }} & {\small{$r=3a$ }} & {\small{$r=4a$ }} & {\small{$r=5a$ }} & {\small{$r=6a$}}\tabularnewline
\hline 
\hline 
{\small{(\ref{eq:BCNW})c }} & \multicolumn{6}{c|}{{\small{8.8(0.5)}}}\tabularnewline
\hline 
{\small{(\ref{eq:BCNW})s }} & {\small{8.8(0.5) }} & {\small{8.7(0.6) }} & {\small{8.6(0.7) }} & {\small{8.6(0.9) }} & {\small{8.8(1.0) }} & {\small{8.9(1.2)}}\tabularnewline
\hline 
{\small{(\ref{eq:1})c }} & \multicolumn{6}{c|}{{\small{7.1(0.6)}}}\tabularnewline
\hline 
{\small{(\ref{eq:1})s }} & {\small{8.6(0.5) }} & {\small{8.2(0.7) }} & {\small{7.7(0.8) }} & {\small{7.3(0.9) }} & {\small{7.0(1.0) }} & {\small{6.7(1.1)}}\tabularnewline
\hline  
\end{tabular}
\par\end{centering}{\small \par}

\caption{\label{TAB003}Results for the topological susceptibility $\chi_{t}a^{4}\times10^{5}$ from fixed topology computations of the static potential ${\mathcal{V}}_{q\bar{q}}(r)$ for various separations. In the column ``method'' the equation number of the expansion is listed, ``c'' denotes a single combined fit for all separations  and ``s'' denotes a separate fit for each separation. As reference value from an unfixed topology computation we use $\chi_{t}a^{4}\times10^{5}=(7.0\pm0.9)$ \cite{de Forcrand:1997sq}.}
\end{table}

Not all of the extracted $\chi_{t}a^{4}$ values perfectly agree with each other or with the result $\chi_{t}a^{4}=7.0\times10^{-5}$ from \cite{de Forcrand:1997sq}, which we take as reference value. There seems to be a slight tension in form of $\approx2\sigma$ discrepancies, when performing fits with the BCNW equation (\ref{eq:BCNW}). The improved version (\ref{eq:1}) gives slightly better results: the majority of the extracted values are less than $1\sigma$ different from the unfixed topology reference value.

One might hope to further improve the results by imposing a stronger constraint, e.g.\ by using only Wilson loops $\langle W_{Q,V}(r,t)\rangle$ with $1/\chi_{t}V,|Q|/\chi_{t}V<0.5$. Indeed there is then consistency with the reference value $\chi_{t}a^{4}=7.0\times10^{-5}$, but the statistical errors are extremely large, of the order of $\chi_{t}a^{4}$ itself or even larger.

We conclude that in principle one can extract the topological susceptibility in Yang-Mills theory from the static potential at fixed topology. In practice, however, one needs rather precise data.

\section{Conclusion}

We have extended equations from the literature \cite{Brower:2003yx,Aoki:2007ka} relating two-point correlation functions at fixed topology to physical hadron masses (i.e.\ hadron masses at unfixed topology). We have also discussed the problem of parity mixing and consequences for the determination of masses of heavier parity partners. Finally we have demonstrated in SU(2) Yang-Mills theory that one can determine the static potential from Wilson loops computed in fixed topological sectors.

\section*{Acknowledgments}

We thank Irais Bautista, Wolfgang Bietenholz, Urs Gerber, H\'{e}ctor Mej\'{\i}a-D\'{\i}az and Christoph P.\ Hofmann for fruitful discussions and collaboration. We also thank Krzysztof Cichy, Dennis Dietrich, Gregorio Herdoiza, Karl Jansen and Andreas Wipf for discussions. We acknowledge support by the Emmy Noether Programme of the DFG (German Research Foundation), grant WA 3000/1-1. This work was supported in part by the Helmholtz International Center for FAIR within the framework of the LOEWE program launched by the State of Hesse.

\end{document}